\documentclass[12pt]{JHEP3}

\usepackage{amsmath,epsfig}
\usepackage{amssymb,amsfonts}
\usepackage{latexsym}
\usepackage[latin1]{inputenc}
\usepackage{xcolor}

\usepackage{graphicx}
\usepackage{longtable}

\relax

\def\be{\begin{equation}}
\def\ee{\end{equation}}
\def\ba{\begin{eqnarray}}
\def\ea{\end{eqnarray}}
\def\D{\Delta}
\def\t{\tau}

\def\D{\Delta}

\def\d{\delta}
\def\nn{\nonumber\\}
\def\pa{\partial}
\def\s{\sigma}

\title{New method for the conformal bootstrap with OPE truncations}

\author{Wenliang Li\\
Department of Physics and Research Institute of Basic Science, Kyung Hee University, Seoul 02447, Korea}

\abstract{ 
We investigate two aspects of conformal field theories. 
In the first part, we study the general 4-point correlator of identical scalars 
around the fully crossing symmetric point $u=v=1$, 
where $u,v$ are conformally invariant cross ratios. 
Since this point is fully crossing invariant, 
we can deduce some general properties of the 4-point correlators from crossing symmetry. 
In the second part, we discuss the conformal bootstrap with OPE truncations. 
As a generalization of Gliozzi's method, 
we propose to extract the low-lying CFT data 
by minimizing the ``error" induced by an OPE truncation. 
The error function $\eta$ measures the violation of crossing symmetry. 
The geometric interpretation of $\eta$ is the length of the vector associated with the truncated OPE. 
As an example, we apply the error-minimization method to the 2d Ising CFT 
with severely truncated OPEs. 
}

\keywords{Conformal Field Theory, Effective Field Theories}

\preprint{}

\begin{document}

\maketitle 

\section{Introduction}
The conformal bootstrap program \cite{Ferrara:1973yt,Polyakov:1974gs,Belavin:1984vu} 
attempts to solve conformal field theories using general principles, 
i.e. conformal symmetry and OPE associativity,  
where OPE stands for Operator Product Expansion. 
In 2d, the conformal symmetry algebra is infinite-dimensional, 
so many 2d CFTs are soluble.  
In higher dimensions, the conformal symmetry algebra is finite-dimensional and less powerful. 
However, since the original work \cite{Rattazzi:2008pe}, 
considerable progress in higher dimensional CFTs has been made using crossing symmetry 
\cite{Rychkov:2009ij}-\cite{Agmon:2017xes}. 
A modern ingredient is the use of efficient numerical methods, 
such as semidefinite programming \cite{Simmons-Duffin:2015qma}. 

The bound method in \cite{Rattazzi:2008pe} is based on a geometric interpretation of the crossing equation, where each conformal multiplet corresponds to a vector and the crossing equation is interpreted as the vanishing of the vector sum associated with the OPE. 
By searching for a separating plane for the potential vectors, 
one can detect the region where no unitary crossing solution can exist and 
rule out the non-unitary parameter space. 
The 3d Ising CFT happens to be located at a special point saturating the bound \cite{ElShowk:2012ht}, 
which leads to precise determinations of the scaling dimensions of the low-lying operators.
 
However, a significant part of the CFT landscape remains unexplored. 
Many interesting CFTs from statistical mechanics are non-unitary and live outside the unitary domain. 
\footnote{In \cite{Hogervorst:2016itc}, the non-unitary logarithmic CFTs are discussed in the context of the conformal bootstrap. }
Even for unitary CFTs, it seems some information is not captured by the bounds. 
Complementary methods for the conformal bootstrap may be useful. 

In general, a geometric vector is characterized by its direction and length. 
The bound method uses the information about the vector directions which are definite due to unitarity. 
Can we also make use of the information about the vector lengths?
Many physical CFTs have a hierarchy in the magnitudes of OPE coefficients. 
The high-lying operators correspond to short vectors due to the suppression by small OPE coefficients. 
\footnote{The high-lying OPE coefficients are generally small if the OPE convergence is rapid.}

In an effective field theory, 
it is typical that infinitely many terms are compatible with the assumed symmetry, 
but we can truncate the effective Lagrangian 
when the coupling constants of the subleading terms are comparably small. 
The effective description is consistent at the energy scales where the presumably leading terms are dominant. 
Analogously, in the context of conformal field theories, 
we should be able to truncate the operator product expansions 
if the subleading OPE coefficients are comparably small. 
These truncated OPEs will be referred to as ``effective OPEs". 
They characterize the effective descriptions of conformal field theories. 
\footnote{This concept can be generalized to quantum field theories without conformal symmetry. 
The Shifman-Vainshtein-Zakharov sum rules \cite{Shifman:1978bx,Shifman:1978by} 
can be considered as an application of effective OPEs in QCD . }

In \cite{Fitzpatrick:2010zm}, the effective conformal field theories were discussed in the context of the AdS/CFT correspondence \cite{Maldacena:1997re,Gubser:1998bc,Witten:1998qj}. 
In contrast to \cite{Fitzpatrick:2010zm}, we do not assume the CFTs under consideration are dual to certain gravitational theories in the bulk. 
Another important difference is that we emphasize the hierarchy in OPE coefficients, instead of operator dimensions \cite{Heemskerk:2009pn}. 
These two hierarchies are related, but not equivalent. 
Operators of higher dimensions usually have smaller OPE coefficients, 
so a hierarchy in operator dimensions may correspond to a hierarchy in OPE coefficients. 
However, operators of low dimensions can decouple and then a hierarchy emerges in OPE coefficients, 
which is sometimes the defining property of physical CFTs.

In \cite{Gliozzi:2013ysa}, Gliozzi proposed an alternative bootstrap method based on OPE truncations. 
\footnote{In fact, OPE truncations are used also in the bound approach, 
as one need to compute low-lying OPE coefficients 
without knowing the scaling dimensions of infinitely many high-lying operators. 
}
In this approach, the operator product expansions are severely truncated 
and the low-lying spectra are captured by the zeros of certain truncation-related determinants. 
\footnote{Let us emphasize that setting these determinants to zero is an approximation. 
A more precise statement would be they are suppressed by the subleading OPE coefficients. 
These determinants will not vanish if we consider the exact spectrum data. } 
This method does not assume unitarity, so it can be applied to non-unitary CFTs 
\cite{Gliozzi:2013ysa}-\cite{Hikami:2017sbg}. 
For instance, the non-unitary Lee-Yang CFTs can be studied using this method. 

Furthermore, based on a severe OPE truncation, 
a new type of universality for different CFTs in various dimensions was found 
using the inverse bootstrap method \cite{Li:2017agi}.  
The low-lying CFT data 
\footnote{The CFT data are the spectral data of the local operators, i.e. scaling dimensions and spins, 
and the OPE coefficients (or the 3-point function coefficients). } 
are consistent with some universal relations 
when the effective OPEs of different CFTs share the same structure. 
This universality suggests effective OPE truncations should have board application. 
\\

In this work, we propose a new method for the conformal bootstrap with OPE truncations. 
The basic idea is to extract the low-lying CFT data by minimizing the ``errors" induced by OPE truncations. 
We will introduce some error functions to quantify the discrepancies 
between crossing symmetry and OPE truncations. 
The error functions are constructed from truncated crossing equations.
The non-trivial
\footnote{There are some trivial minima corresponding to unphysical solutions.}
minimum of an error function can vanish, 
when the number of equations in the error function 
coincides with the number of free parameters in the truncated OPE. 
In this case, the error-minimization method is equivalent to Gliozzi's determinant method. 
As a generalization of Gliozzi's method, 
we consider error functions involving more equations. 
The nontrivial minima become strictly positive and 
the predictions are different from those of Gliozzi's method. 
We expect to obtain more accurate results by increasing 
the number of equations in the error functions. 

In \cite{ElShowk:2012hu}, an analogous minimization procedure was introduced. 
After the scaling dimensions of low-lying operators were extracted from the extremal functional, 
the minimization procedure was performed to determine the OPE coefficients. 
In some sense, the new method proposed in this work also generalizes the minimization procedure in \cite{ElShowk:2012hu}. 
The difference is that we minimize the errors to extract both the spectral information and OPE coefficients. 
As a result, we do not need to assume unitarity which is crucial to the extremal functional method in \cite{ElShowk:2012hu}. 

Let us emphasize that our ``errors" are not the standard errors 
that measure the accuracy of approximate CFT data. 
These error functions are devised to measure the violations of crossing symmetry due to OPE truncations, 
or the accuracy of the truncated crossing equations.  
The geometric interpretation of an error function is the vector length associated with a truncated OPE. 
\footnote{If an OPE is exact, the vector length should vanish 
because the complete OPE corresponds to a zero vector. 
Furthermore, the vector length corresponding to a truncated OPE should be small when the OPE convergence is fast. } 
The assumption behind the new method is that 
some physical OPEs converge rapidly and their truncated counterparts correspond to the minima of the error function. 
\footnote{In \cite{El-Showk:2014dwa}, it was conjectured that the 3d Ising CFT minimizes the central charge 
in the landscape of 3d unitary CFTs. 
In this work, we are proposing some physical CFTs minimize the error function. } 

Admittedly, it is not yet clear how to set the rigorous error bars, 
but we can examine the stability of the predictions by increasing the number of equations, 
which provides a criterion for selecting sensible results. 
Then we can estimate the errors in the predictions, based on the fact that crossing equations are only approximately solved. 
\\

The structure of this paper is as follows. 
In section \ref{fullcrossing}, we discuss the general 4-point correlator of identical scalars 
and the conformal bootstrap 
around the fully crossing symmetric point $u=v=1$. 
In section \ref{mini-error}, we present the method of error minimizations 
for the conformal bootstrap with OPE truncations. 
In section \ref{ex-Ising}, we apply the error-minimization method to the 2d Ising CFT. 
In section \ref{discussion}, we summarize our results and discuss future directions. 

\section{Fully crossing symmetric point $u=v=1$}\label{fullcrossing}
In this section, we will discuss the fully crossing symmetric point
\be
u=v=1,\label{CIP}
\ee
where the cross ratios $u, v$ are invariant under all crossing transformations. 
The conformal bootstrap around this crossing symmetric point has not been systematically investigated, 
in contrast to the standard one $u=v=1/4$. 
\footnote{
From the general analysis in \cite{Pappadopulo:2012jk}, 
in unitary CFTs, 
the OPE convergence of the 4-point correlators around $u=v=1/4$ is faster than that around $u=v=1$. 
But the prefactor of the estimate in \cite{Pappadopulo:2012jk} is suboptimal, which was explained in \cite{Rychkov:2015lca}. 
See also appendix A of \cite{Rychkov:2016mrc} for a concrete example in the 3d Ising CFT. 
}
But this special point \eqref{CIP} has intrinsic value for the conformal bootstrap as it is fully crossing invariant. 
In section \ref{Gen-crossing-sol}, we will study the general crossing solution expanded around $u=v=1$, 
which takes the form of a standard Taylor series. 
\footnote{The general crossing solution in terms of this standard basis may be useful for the inverse approach of the conformal bootstrap \cite{Li:2017agi}.}
We will also discuss how to characterize the functional deviations of an interacting correlator from the correlator of generalized free fields. 

In section \ref{full-crossing-bootstrap}, we will formulate the conformal bootstrap equations around $u=v=1$ 
and fix some notations to be used in section \ref{mini-error}. 
When defining the error functions in section \ref{mini-error}, we will treat equally the equations from different numbers of derivatives. 
Due to this naive construction, we will not consider the expansion around $u=v=1/4$. 
Instead, the equations from the expansion around $u=v=1$ are more appropriate. 
We will come back to this point in section \ref{mini-error}.
In addition, an advantage of expanding around $u=v=1$ is that 
we only need a small number of series coefficients of conformal blocks 
to achieve preliminary numerical precision. 
\footnote{The series representation of the conformal block related to a spin-$l$ primary operator reads
\be
F_{\t,l}(u,v)=\sum_{m,n=0}^\infty b_{m,n}(d,\t,l)\, u^{\t/2+m}(1-v)^n,
\ee
where $d$ is the spacetime dimension 
and $\t,l$ are the twist and spin of the exchanged primary. 
The series coefficients $b_{m,n}$ are rational function of $\t,l,d$, 
which are determined by the Casimir equation with the boundary condition $b_{0,l}=1$ \cite{Dolan:2000ut, Dolan:2003hv}. See appendix A of \cite{Li:2017agi} for more details. 
To obtain the numerical results in section \ref{ex-Ising}, we expand $u^{\t/2+m}$ around $u=1$. 
For our precision, we only need to know $b_{m,n}(d,\t,l)\big|_{d\rightarrow 2}$ with $m=0,1,\dots, 10$. 
The leading digits remain the same if we increase the maximum of $m$. 
}
As a result, when the error-minimization method is explained in section \ref{mini-error}, 
we will consider the conformal bootstrap around $u=v=1$. 
Note that it is not necessary to formulate the error-minimization method around this point. 

\subsection{General crossing solution}\label{Gen-crossing-sol}
The conformally invariant cross ratios $u,v$ of 4-point correlators are defined as 
\be
u=\frac {x_{12}^2 x_{34}^2}  {x_{13}^2 x_{24}^2}
,\quad
v=\frac {x_{14}^2 x_{23}^2}  {x_{13}^2 x_{24}^2},
\ee
where $x_{ij}=|x_i-x_j|$, and $x_i$ denotes the position of operator $\phi_i$. 
Under crossing transformations, they transform as
\ba
1\leftrightarrow 3&:&\quad u\to v,\quad v\to u,\label{crossing-1}\\
1\leftrightarrow 2&:&\quad u\to u/v,\quad v\to 1/v,\label{crossing-2}\\
1\leftrightarrow 4&:&\quad u\to 1/u,\quad v\to v/u.\label{crossing-3}
\ea
The line $u=v$ is invariant under the first crossing transformation \eqref{crossing-1}, 
while the special point $u=v=1$ is invariant under all the crossing transformations \eqref{crossing-1}, \eqref{crossing-2}, \eqref{crossing-3}. 
In terms of the variables $z,\,\bar z$, which are defined by $u=z\bar z$ and $v=(1-z)(1-\bar z)$, 
the fully crossing symmetric point is at $z=1/2+ i \sqrt 3/2,\,\bar z=1/2- i \sqrt 3/2$. 

Let us consider the 4-point function of identical scalars, 
\be
< \phi(x_1)\,\phi(x_2)\,\phi(x_3)\,\phi(x_4)>
=\frac 1 {(x_{12}^2 x_{34}^2)^{\D}}\, G(u,v),
\ee
where $\D$ is the scaling dimension of $\phi$.

Invariance under the first crossing transformation \eqref{crossing-1} implies
\be
u^{-\D} G(u,v)=v^{-\D} G(v,u). \label{G1}
\ee

Let us introduce $H(u,v)=u^{-\D} G(u,v)$. 
From the first crossing equation \eqref{G1}, $H(u,v)$ is a symmetric function
\footnote{An equivalent statement is that $v^{\D} G(u,v)$ is a symmetric function, 
which was used as the definition of $H(u,v)$ in \cite{Li:2017agi}. 
In this work, we change the definition of $H(u,v)$ for the exact solution to be discussed below. 
}
\be
H(u,v)=H(v,u).\label{crossing-1-H}
\ee

The symmetric function $H(u,v)$ expanded around $u=v=1$ is a Taylor series of two variables
\be
H(u,v)=\sum_{m,n=0}^\infty c_{m,n}(u-1)^m (v-1)^n,
\quad\text{with}\quad
c_{m,n}=c_{n,m}.
\ee
Since $c_{m,n}$ is symmetric, let us consider $c_{m,n}$ with $m\geq n$ as the independent parameters. 
Under the second crossing transformation \eqref{crossing-2}, invariance of the 4-point correlator indicates
\be
G(u,v)=G(u/v,1/v).\label{G2}
\ee
The second crossing equation \eqref{G2} imposes further constraints on the symmetric function $H(u,v)$
\be
H(u,v)=u^{-\D}G(u/v,1/v)=v^{-\D}H(u/v,1/v),\label{crossing-2-H}
\ee
which leads constraints on the series coefficients $c_{m,n}$. 
The third crossing transformation \eqref{crossing-3} does not lead to an independent equation. 

The second crossing equation \eqref{crossing-2-H} can be solved order by order
\be
H(u,v)=\sum_{k=0}^\infty H^{(k)}(u,v),
\ee
where $k$ indicates the order
\be
H^{(k)}(u,v)=\sum_{m=0}^k c_{m,\,k-m}\,(u-1)^{m}(v-1)^{k-m}.
\ee 
A few examples at low orders are:
\begin{itemize}
\item 
For $k=0$,
\be
H^{(0)}(u,v)=c_{0,0}\,,
\ee
where $c_{0,0}$ is a free parameter.  
The arbitrariness of $c_{0,0}$ reflects the fact that $H(u,v)$ is crossing invariant at $u=v=1$. 
\item
For $k=1$,
\be
H^{(1)}(u,v)=c_{1,0}(u-1)+c_{1,0}(v-1),
\ee
where $c_{1,0}$ is related to $c_{0,0}$ as
\be
c_{1,0}=-\frac 1 3 \D\, c_{0,0}.\label{k1}
\ee
There is no independent parameter at this order. 
\item
For $k=2$,
\be
H^{(2)}(u,v)=c_{2,0}(u-1)^2+c_{1,1}(u-1)(v-1)+c_{2,0}(v-1)^2,
\ee
where $c_{1,1}$ is a free parameter and $c_{2,0}$ is given by
\be
c_{2,0}=\frac 1 {3\times 2!} (\D)_2\, c_{0,0}-c_{1,1}\,. \label{k2}
\ee
The notation $(x)_n$ is the Pochhammer symbol
\be
(x)_n=\Gamma(x+n)/\Gamma(x).
\ee
One may consider $c_{2,0}$ as an arbitrary parameter and express $c_{1,1}$ in terms of $c_{0,0}$ and $c_{2,0}$. 
However, from the results at higher orders, 
it is more natural to consider the coefficients with $m\sim n$ as the independent parameters.
\item
For $k=3$,
\be
H^{(3)}(u,v)=\sum_{m=0}^3 c_{m,\,3-m}\,(u-1)^{m}(v-1)^{3-m},
\ee 
where $c_{3,0}$ is not independent and we have
\be
c_{3,0}=-\frac 1 {3\times 3!} (\D)_3\, c_{0,0}
+\frac 1 3 (\D+2)\,c_{1,1}
-\frac 2 3\, c_{2,1}\,.\label{k3}
\ee
According to the comment in the $k=2$ case, 
we consider $c_{2,1}$ as the new independent parameter at this order. 
\end{itemize}

It is straightforward to solve the second crossing equation \eqref{crossing-2-H} to higher orders. 
To the order $k=6$, $H(u,v)$ has 28 coefficients as a generic Taylor series. 
Using the first crossing equation \eqref{crossing-1-H}, only 16 coefficients are independent. 
Then, from the second crossing equation \eqref{crossing-2-H}, there remain 7 independent parameters:  
$c_{0,0}$, $c_{1,1}$, $c_{2,1}$, $c_{2,2}$, $c_{3,2}$, $c_{3,3}$, $c_{4,2}$. 
Note that we have two independent parameters at the order $k=6$. 
\\

We notice that $c_{0,0}$ appears only in $c_{n,0}$ 
and the coefficients of $c_{0,0}$ have a simple closed form expression
\be
c_{n,0}=(-1)^n \frac {(\D)_n} {3 n!}\, c_{0,\,0}+\dots,\quad n>0\,.
\ee
Let us consider the limit where the cross terms vanish. 
Then all the non-zero coefficients are proportional to $c_{0,0}$ and the complete crossing solution becomes
\be
c_{0,0}+\sum_{n=1}^\infty (-1)^n \frac {(\D)_n} {3 n!}\, c_{0,\,0}
\left[
(u-1)^n+(v-1)^n
\right]
=
\frac {c_{0,0}}3\,(1+u^{-\D}+v^{-\D}),
\ee
which is a rescaled version of the 4-point function of the fundamental scalar in a generalized free theory. 
\footnote{In a generalized free theory, 
the 4-point functions of composite operators have different dependence on $u,v$. 
We focus on the correlator of the fundamental scalar. }
The normalization is fixed by $c_{0,0}$. 

If the coefficients of the cross terms are small, 
the 4-point correlator around $u=v=1$ can be considered 
as a small deformation of the generalized free correlator. 
This then leads to the issue of how to characterize the deviation from the generalized free theory. 
\\

In a generic CFT, 2-point and 3-point correlators have the same functional forms 
as the generalized free ones due to conformal symmetry, 
but they can have different normalizations related to the OPE coefficients. 
In contrast, the functional form of a conformal 4-point correlator is not fully determined by conformal symmetry. 

Since $u=v=1$ is invariant under all the crossing transformations, 
crossing symmetry is manifest around this fully crossing symmetric point. 
With this additional symmetry, we can deduce some general properties of the 4-point correlators. 
The zeroth order coefficient $c_{0,0}=H(1,1)$ corresponds to the normalization factor, 
which is analogous to 3-point function coefficients. 
The first order coefficients $c_{0,1}, \, c_{1,0}$ are fixed by crossing symmetry, 
which means they coincide with those of the generalized free correlator, up to the normalization determined by $c_{0,0}$. 
In this sense, assuming $H(1,1)\neq 0$, all 4-point correlators are weakly interacting around the fully crossing symmetric point $u=v=1$. 
\footnote{
An example of $H(1,1)=0$ is the 4-point function of the energy operator in the 2d Ising CFT
\ba
H(u,v)&=&1 - u - v + u^2 - u v + v^2
\nn&=&
(u-1)^2+(v-1)^2-(u-1)(v-1)+\dots,
\ea
where the zeroth and first order terms vanish.
}

Deviations from the generalized free theory start at the second order. 
In the generalized free correlator, there is no cross term, so the coefficient $c_{1,1}$ vanishes. 
We propose to use the ratio $R=c_{1,1}/c_{0,0}$ to characterize the leading deviation around $u=v=1$. 
By construction, this ratio $R$ vanishes in the generalized free theory. 
We can compute the exact values of $R$ in some 2d CFTs. 
In the 2d Ising CFT, the 4-point correlator of the spin correlator gives $R_{\text{2d Ising}}=-7/576\sim -0.012$. 
\footnote{For the 3d Ising CFT, a rough estimate gives $R_{\text{3d Ising}}\sim 0.05$.}
In the 2d Lee-Yang CFT, the 4-point correlator of the lowest scalar gives $R_{\text{2d LY}}\sim 0.025$. 
It is also natural to introduce higher order analogues $R_{m,n}=c_{m,n}/c_{0,0}$ using the other independent parameters. 

The deviations from the generalized free theory was also discussed in \cite{Rychkov:2016mrc}, which was referred to as ``non-Gaussianity". 
Let us compare the measure of non-Gaussianity in \cite{Rychkov:2016mrc} with our discussion above. 
In \cite{Rychkov:2016mrc}, the non-Gaussianity is measured by $Q$, 
the ratio of the interacting and generalized free correlators, 
as a function of the cross ratios. 
According to the criterion in \cite{Rychkov:2016mrc}, 
the spin correlators of the Ising CFTs in 2d and 3d show significant non-Gaussianity 
at $u=v=1$, 
because they correspond to the minima of $Q$ in the considered domain. 
This is different from our measure. 
Using a different ratio, we consider the functional deviations, which is generally weak around $u=v=1$. 
For instance, this is reflected by the plateau behavior of the Ising correlators shown in \cite{Rychkov:2016mrc}. 
In addition, the reference point $z=0$ in \cite{Rychkov:2016mrc} shows strong deviations from our functional perspective, 
which can be seen from the peak behavior \footnote{The $Q$-surface at $z=0$ has a conical singularity.}. 
Nevertheless, the two ratios $Q, R$ are related, because $c_{1,1}=R\, H(1,1)$ measures 
the curvature of the $Q$-surface at $u=v=1$. 
From both perspectives, the Ising CFTs show significant departure from the generalized free theory in certain region of the Euclidean regime.   
\\

An interesting observation of the general crossing solution around $u=v=1$ is that the $\D$-dependence can be separated. 
To explain this observation, let us define $\bar H^{(k)}(u,v)$ as
\be
H(u,v)=\sum_{k=0}^\infty (\D)_k\,\bar H^{(k)}(u,v),
\ee
with
\be
\bar H^{(k)}(u,v)=\sum_{m=0}^k \bar c_{m,\,k-m}\,(u-1)^{m}(v-1)^{k-m}.
\ee 

Then \eqref{k1}, \eqref{k2}, \eqref{k3} become $\D$-independent equations:
\be
\bar c_{1,0}=\frac 1 3 \bar c_{0,0},
\quad
\bar c_{2,0}=\frac 1 {3\times 2!} \bar c_{0,0}-\bar c_{1,1},
\quad
\bar c_{3,0}=- \frac 1 {3\times 3!}\bar c_{0,0}+\frac 1 3 \bar c_{1,1}-\frac 2 3 \bar c_{2,1}.
\ee
The absence of $\D$ is verified by the explicit solutions of $\bar c_{m,n}$ to the order $k=6$. 
There may be more interesting results about the general structure of crossing solutions around $u=v=1$. 

\subsection{Conformal bootstrap}\label{full-crossing-bootstrap}
According to operator product expansion, 
the conformal invariant part of the 4-point function is decomposed into conformal blocks
\be
G(u,v)=1+\sum_{i} P_i \,F_{\t_i,\, l_i}(u,v),\label{OPE}
\ee
where $i$ labels the exchanged primary operators, 
$P_i$ are the squared OPE coefficients and 
$F_{\t_i,\, l_i}$ are the conformal blocks. 

Crossing symmetry implies non-trivial equations for the CFT data. 
According to the conformal block expansion \eqref{OPE}, 
the first crossing equation \eqref{G1} expanded around $u=v=1$ gives
\be
(\pa_u^m\pa_v^n-\pa_u^n\pa_v^m)(u^{-\D})\Big|_{u=v=1}+\sum_{i} P_i\, (\pa_u^m \pa_v^n-\pa_u^n \pa_v^m)\left[u^{-\D}F_{\t_i,\, l_i}(u,v)\right]\Big|_{u=v=1}
=0. 
\ee
The independent equations are
\be
(-1)^m (\D)_m\, \d_{n,0}+\sum_i P_i\, a_{m,n}(\D,\,\t_i,\,l_i)=0,\quad m>n,
\label{crossing-eq-expanded}
\ee
where $a_{m,n}$ are defined as
\be
a_{m,n}(\D,\,\t_i,\,l_i)=(\pa_u^m \pa_v^n-\pa_u^n \pa_v^m)
\left[u^{-\D}F_{\t_i,\, l_i}(u,v)\right]\Big|_{u=v=1}.\label{def-amn}
\ee
When a primary operator has even spins, 
the associated conformal block satisfies the second crossing equation \eqref{G2}. 
Therefore, the second crossing equation \eqref{G2} is solved when all the exchanged primaries have even spins. 

\section{Minimizing the errors from OPE truncations}\label{mini-error}
In this section, we will present a new method for the conformal bootstrap 
where operator product expansions are truncated. 
An operator product expansion is usually complicated, 
as the expansion leads to infinitely many operators in the sum. 
We will focus on the case where the OPE can be effectively truncated, 
which means the contributions from the high-lying operators are subleading. 
An effective OPE has finitely many parameters, 
so the bootstrap problem is more approachable. 
However, crossing symmetry is broken by the OPE truncation. 
We expect the crossing violation is weak 
when the low-lying CFT data are close to those of the effective OPEs. 
Therefore, we can extract the low-lying CFT data 
by minimizing the violation of crossing symmetry due to the OPE truncation. 
We will measure the crossing violation by some ``error" functions, 
so the new method will be referred to as the error-minimization method. 

The crossing equations \eqref{crossing-eq-expanded} can be rewritten as
\be
(-1)^m (\D)_m\, \d_{n,0}+\sum_{i=1}^N P_i\, a_{m,n}(\D,\,\t_i,\,l_i)
=-\sum_{i>N} P_i\, a_{m,n}(\D,\,\t_i,\,l_i),
\label{trun-crossing}
\ee
where $N$ is the number of exchanged primaries in the truncated OPE
\footnote{In this schematic form, we do not write explicitly the coordinate dependence and the descendant operators, 
as they are fully determined by conformal symmetry. 
In the precise expression of \eqref{trun-OPE}, 
$P_i$ should be replaced by $C_{\phi \phi \mathcal O_i}$, the OPE coefficients. 
In \eqref{trun-OPE}, we use squared OPE coefficients $P_i$ 
because they are more relevant to the conformal block expansions in this work. 
} 
\be
\phi\times\phi=I+\sum_{i=1}^N P_i\, \mathcal O_i,\label{trun-OPE}
\ee
and the right hand side of \eqref{trun-crossing} 
is related to the high-lying operators omitted in the OPE truncation. 
Let us focus on the case $n=0$, 
so the first term on the left hand side of \eqref{trun-crossing} is present 
and the convergence is usually rapid. 
\footnote{If a correlator has a weak deviation from the generalized free correlator around $u=v=1$, 
the coefficients of the cross terms ($n>0$) are small 
and require more operators to converge in the OPE.}

Before an exact solution is available, 
we do not know the exact content omitted in an OPE truncation, 
which are encoded in the right hand side of \eqref{trun-crossing}. 
However, according to \eqref{trun-crossing}, 
the error induced by the OPE truncation is equal to the left hand side of \eqref{trun-crossing}. 
Therefore, we should optimize the low-lying CFT data 
such that the left hand side of \eqref{trun-crossing} is close to zero, 
which means the OPE-truncation error is minimized . 
\\

Let us define the ``error" induced by an OPE truncation as
\footnote{There are many possible ways to quantify the OPE-truncation error. 
Here we use one of the simplest definitions.}
\be
\eta(k)=\sqrt{\frac 1 k \sum_{m=1}^k \left[f_{m}(\D,\,\t_i,\,l_i)\right]^2},\label{error}
\ee
where $f_{m}$ are the truncated crossing equations, i.e. the left hand side of \eqref{trun-crossing} with $n=0$, 
\be
f_{m}(\D,\,\t_i,\,l_i)=(-1)^m (\D)_m+\sum_{i=1}^N P_i\, a_{m,0}(\D,\,\t_i,\,l_i), \label{trun-crossing-eq}
\ee
and $k$ is the number of $f_m$ in the error function $\eta(k)$. 
These error functions measure the violations of crossing symmetry due to OPE truncations. 
We want to extract the physical CFT data by minimizing $\eta(k)$. 

In \eqref{error}, we use the equations from the expansion around $u=v=1$. 
Let us emphasize that the OPE convergence rate of a 4-point correlator at a point is different from 
that of its Taylor coefficients around the same point. 
Around $u=v=1/4$, the OPE convergence becomes less rapid when we increase the number of u- or v- derivatives.
\footnote{See the appendix of \cite{El-Showk:2016mxr} for a discussion about the OPE convergence rate of derivatives of the 4-point correlator. } 
This is related to the fact that $u=v=1/4$ is on the boundary of the Euclidean regime. 
The equations from a large number of derivatives are probing the Lorentzian regime where the OPE convergence is slow. 
\footnote{For s-channel OPE of a 4-point correlator, the convergence rate decreases significantly as we introduce more v-derivatives. }
In contrast, for $u=v=1$, the OPE convergence is less sensitive to the number of derivatives, 
because this point is far from the Lorentzian regime. 
To make use of the rapid convergence around $u=v=1/4$, 
we need to introduce more sophisticated error functions, which will not be covered in this work.
\\

The error function \eqref{error} has a geometric interpretation. 
We can think of the conformal block of a conformal multiplet $\{\mathcal O_i,\,\pa^k \mathcal O_i\}$ as a vector
\be
\vec F_{(i)}=\{a_{m,n}(\D,\,\t_i,\,l_i)\},\quad m>n\geq 0,
\ee
whose components are labeled by two integers $m,n$. 
\footnote{According to \eqref{def-amn}, $a_{m,n}$ is antisymmetric in $m,n$. 
The symmetric components of the vector are not written explicitly because they are irrelevant to the crossing equations. }
Then the exact crossing equation \eqref{crossing-eq-expanded} indicates the infinite vector sum below gives the zero vector:
\be
\vec I+\sum_{i=1}^\infty P_i\,\vec F_{(i)}=\vec 0\label{crossing-vec},
\ee
where the identity operator is related to the vector $\vec I$ with
\be
\vec I=\{(-1)^m (\D)_m\,\d_{0,n}\},\quad m>n\geq 0.
\ee
To have the notion of length, 
let us introduce some inner products 
\footnote{Strictly speaking, we do not need to introduce an inner product to define a norm. 
In addition, some norms are not given by inner products, 
such as the p-norms with $p\neq 2$. 
In \eqref{error}, we use the 2-norm, 
but the minimization in \cite{ElShowk:2012hu} is based on the 1-norm \cite{Behan:2017rca}.}
\be
<\vec A,\,\vec B>_{(k)}\,=\frac 1 k \sum_{m=1}^k A_{m,0}\, B_{m,0}. \label{innerprod}
\ee
We can also define angles using the inner products. 
\footnote{The linear functional in the bound method can be interpreted as the inner product with a fixed vector. 
If there exists a special vector such that its inner products with the conformal block vectors have a definite sign,  
then no unitary crossing solution is possible. 
Note that the associated inner product is different from \eqref{innerprod}. 
}
According to the definitions in \eqref{innerprod}, 
the error functions $\eta(k)$ coincide with the norms 
of the vector sum corresponding to the truncated OPE \eqref{trun-OPE}
\be
\eta(k)=\left\Vert\vec I+\sum_{i=1}^N P_i\,\vec F_{(i)}\right\Vert_{(k)}. 
\ee
In general, we can reverse the procedure and define the error functions as the vector norms associated with the truncated OPEs. 
If the CFT data of the truncated OPE are exact, 
an error function also gives the norm of the vector sum 
corresponding to the high-lying operators
\be
\eta(k)\Big|_{\text{exact}}=\left\Vert\sum_{i=N+1}^\infty P_i\,\vec F_{(i)}\right\Vert_{(k)},\label{eta-highlying}
\ee
where \eqref{crossing-vec} is used. 
\\

Let us discuss the relation between Gliozzi's determinant method \cite{Gliozzi:2013ysa} and our error-minimization method. 
When $k$ equals 
\footnote{When $k$ is smaller than the number of free parameters, there are continuous families of unphysical solutions of $\eta=0$.} 
the number of free parameters in the truncated OPE, 
it is sometimes possible to minimize $\eta(k)$ to zero. 
If so, there exists a set of truncated CFT data 
which solve $k$ truncated crossing equations \eqref{trun-crossing-eq}. 
We can interpret these equations as a system of linear equations, 
where the unknowns are the squared OPE coefficients $P_i$ 
and the coefficients of the unknowns are $a_{m,0}$ constructed from conformal blocks. 
To have a solution, at most $k_P$ equations are linearly independent. 
Note that $k_P$ is the number of unknown OPE coefficients and 
$k_\t=k-k_P$ is the number of unknown twists. 
\footnote{We assume only totally symmetric fields are present. 
We use twists and spins as the independent spectral data. 
The scaling dimension of a spin-$l$ operator with twist-$\t$ is $\D=\t+l$.}
Let us imagine the constant part $(-1)^m (\D)_m+\dots$ is multiplied by an unknown whose solution is $P=1$, 
where $\dots$ indicates the terms with known OPE coefficients. 
Then we can construct $(k_P+1)$ by $(k_P+1)$ matrices from the coefficients of $(k_P+1)$ equations. 
Their determinants should vanish, 
otherwise the systems of $(k_P+1)$ homogeneous equations do not have any solution. 
In short, any set of $(k_p+1)$ equations should be linearly dependent.  
A solution with a vanishing ``error" is precisely a crossing point of the hypersurfaces with vanishing determinants in Gliozzi's method. 
\footnote{In \cite{Gliozzi:2013ysa}, the expanded function is normalized by the identity contribution, 
so only one equation is non-homogeneous. 
Although they may look different, 
the determinants from our construction are equivalent to Gliozzi's determinants 
if we also expand around $u=v=1/4$. 
The reason is that the independent equations do not depends on how we normalize the expanded function. 
After we change the normalization, 
the equations from the expansion are linear combinations of those before the normalization change.}

When the OPE convergence is rapid, 
the contributions of the high-lying operators are subleading, 
then the truncated crossing equations \eqref{trun-crossing-eq} 
from the same truncated OPE have weak linear independence. 
But they do not have exact linear dependence and the corresponding determinants should not vanish. 
As we increase $k$, 
the number of equations becomes greater than that of free parameters, 
then it is usually impossible to solve all the truncated equations \eqref{trun-crossing-eq} at the same time. 
The error function $\eta(k)$ becomes strictly positive. 
\footnote{We assume the squared OPE coefficients are real numbers. 
In other words, we assume the OPE coefficients are either real or purely imaginary.  } 
From \eqref{eta-highlying}, we expect the magnitude of the error function $\eta$ at the minimum 
is related to $P_{N+1}$
\be
\eta\sim |P_{N+1}|,
\ee 
where $P_{N+1}$ is the high-lying squared OPE coefficient with the largest absolute value. 
The precise coefficient in front of $|P_{N+1}|$ depends on the definition of $\eta$, 
the CFT under consideration
\footnote{For example, degeneracy or density of the states can introduce a multiplicative factor. }
, the truncation scheme and the normalization convention of the conformal blocks 
\footnote{Our normalization convention of conformal blocks is: $F_{\t,l}(u,v)=u^{\t/2}(1-v)^l+\dots$ .}.

\section{Example: the 2d Ising CFT}\label{ex-Ising}
In this section, we will apply the error-minimization method to the 2d Ising CFT. 
We mainly consider a severely truncated OPE
\be
\s\times\s=I+P_\epsilon\,\epsilon+P_T\,T,\label{OPEtrun}
\ee
where $\s$ is the spin operator, 
$I$ is the identity operator, 
$\epsilon$ is the energy operator, 
$T$ is the stress tensor, 
and $P_\epsilon,\,P_T$ are the squared OPE coefficients of $\epsilon, T$. 
\footnote{In the precise expression, we should substitute $P_\epsilon,\,P_T$ with the OPE coefficients $C_{\s\s\mathcal \epsilon},\,C_{\s\s T}$.}
\\

Let us assume the scaling dimension $\D_\s$ of the spin operator $\s$ takes the exact value
\be
\D_\s=1/8,
\ee
which is equal to the twist $\t_\s$ because $\s$ is a scalar.  
Then we consider the error function $\eta(k)$ with various $k$ and 
extract the low-lying CFT data by minimizing $\eta(k)$. 
Note that $k$ is the number of truncated crossing equations \eqref{trun-crossing-eq}  in $\eta(k)$. 
In Table \ref{exact-Delta-sigma}, 
we summarize the predictions of the low-lying CFT data from the error-minimization method. 
We extract the results from the local minima in the domain of $\D_\epsilon>1/2$. 
\footnote{The global minimum is located at $\D_\epsilon=0,\,P_\epsilon=-1,\,P_T=0$ with $\eta(k)=0$, 
which corresponds to the unphysical solution $G(u,v)=0$. }
We can see the approximate CFT data become more accurate as we increase $k$ 
and then stay around the exact values with errors of order $10^{-3}$.  
The minima of $\eta(k)$ with $k>3$ are of order $10^{-4}$, 
which are in accord with the magnitudes of the OPE coefficients of the leading high-lying operators, 
the level-4 Virasoro descendants. 
When $k>4$, the minimum of $\eta(k)$ is smaller for a larger $k$. 

\begin{table}[h!]
\centering
 \begin{tabular}{||c ||c c c c c c c c|| c ||} 
 \hline
 	$k$		& 3 		&  4 & 5 & 6 & 7 & 8 & 9 & 10  & exact\\ [0.5ex] 
 \hline
 \hline
$\D_\epsilon$ 	& 1.063 	& 1.021	&1.008 &1.001	&0.997 &0.996	&0.995	&0.996 & 1 \\
\hline
$P_\epsilon$ 	& 0.233 	& 0.242	&0.245&0.247	&0.248	&0.248	&0.248	&0.248 &0.25 	\\\hline
$P_T$ 		& 0.015 	& 0.015	&0.014 &0.014 &0.014 &0.014 &0.014 &0.014  &0.016	\\\hline
\end{tabular}
\caption{The predictions from the error-minimization method of the low-lying CFT data of the 2d Ising CFT and the corresponding exact values. 
We use a severely truncated OPE $\s\times\s=I+P_\epsilon\,\epsilon+P_T\,T$ and assume $\D_\s=1/8$.  
$\D_\s,\,\D_\epsilon$ are the scaling dimensions of the spin and energy operators. 
$P_\epsilon,\, P_T$ are the squared OPE coefficients of the energy operator and the stress tensor. 
As we increase $k$, the predictions of $\D_\epsilon, P_\epsilon$ become more accurate: 
the differences between the predictions and exact values are reduced by one order of magnitude. }
\label{exact-Delta-sigma}
\end{table}

Now we estimate the errors in the stable predictions. 
The predictions contain errors because the first crossing equation \eqref{G1} is only approximately solved, 
which is related to the fact that $\eta(k)$ is non-zero. 
To set the error bars, let us examine the truncated crossing equations \eqref{trun-crossing-eq} one by one. 
When estimating the error in a CFT parameter, we consider this parameter as an unknown quantity and 
determine it from the truncated crossing equation $f_m=0$ using the other approximated CFT data. 
In this way, the OPE-truncation error in the crossing equation is all absorbed into this CFT parameter, 
so it will be different from the prediction of the error minimization. 
The result depends on which $f_m$ is used and the error bar will be set by the largest deviation from the error-minimized value. 

In the case of $k=10$, the predictions from the error-minimization method read
\be
k=10:\quad\D_\epsilon=0.9955(88),\quad
P_\epsilon=0.2484(22),\quad
P_T=0.0142(13),
\ee
which are compatible with the exact values
\be
\D^{\text{exact}}_\epsilon=1,\quad
P^{\text{exact}}_\epsilon=0.25,\quad
P^{\text{exact}}_T=1/64\sim 0.0156.
\ee
The error in $P_T$ is slightly underestimated. 
As an example, let us explain in more detail how the error bar of $\D_\epsilon$ is determined. 
We solve $\D_\epsilon$ from $f_m=0$ with $m=1,2,\dots, 10$, using $P_\epsilon=0.2484$ and $P_T= 0.0142$. 
The results are around $\D_\epsilon= 0.9955$, but with certain deviations. 
The largest deviation is from $f_3=0$, which gives $\D^{m=3}_\epsilon= 1.0043$. 
The error bar is obtained from the difference $1.0043-0.9955=0.0088$. 
\\

\begin{figure}[h!]
\begin{center}
\includegraphics[width=14cm]
{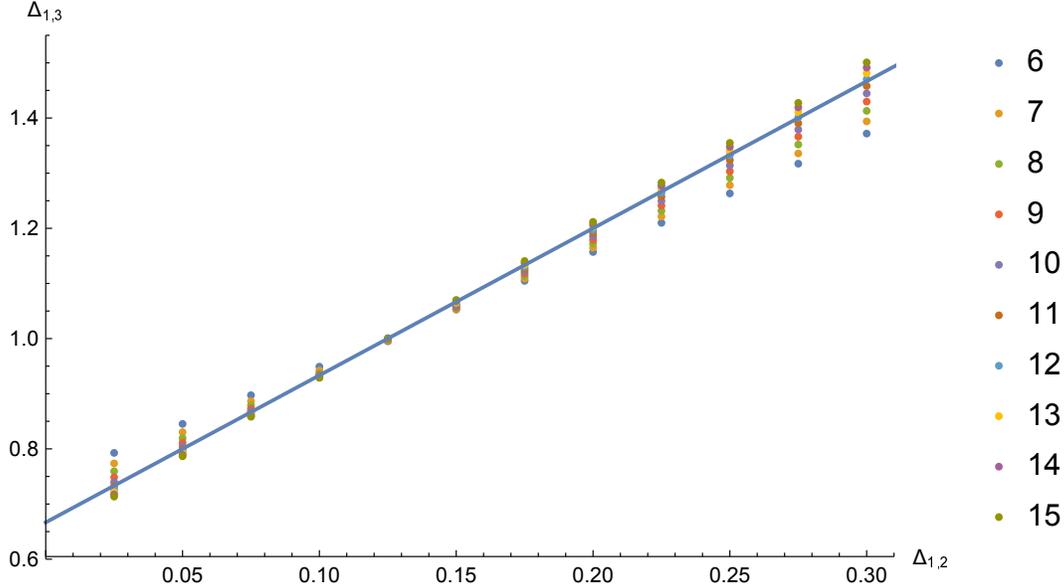}
\caption{The exact relation (solid line) between $\D_{1,2}$ and $\D_{1,3}$, 
and the predictions (points) of $\D_{1,3}$ from the error-minimization method. 
We consider the truncated OPE: 
$\phi_{1,2}\times\phi_{1,2}=I+P_{1,3}\,\phi_{1,3}+P_T\,T$. 
We use the error function $\eta(k)$ with $k=6, 7, \dots,15$. 
We choose $\D_{1,2}=0.025 j$ with $j=1,2,\dots, 12$. 
The predictions at $\D_{1,2}=0.125$ are very stable and match the exact value. 
The predictions become less accurate and less stable when $\D_{1,2}$ is far from $0.125$. 
When $\D_{1,2}<0.125$, the highest point corresponds to $k=6$ 
and the prediction decreases as we consider a larger $k$. 
When $\D_{1,2}>0.125$, the lowest point is associated with $k=6$ 
and the result increases as we choose a larger $k$. 
} 
\label{fig-near-Ising}
\end{center}
\end{figure}

Then we relax the assumption about the scaling dimension $\D_\s$ of the scalar operator $\s$. 
The spin and energy operators $\s,\,\epsilon$ are generalized to 
the Virasoro primaries $\phi_{1,2},\,\phi_{1,3}$ in the 2d CFTs. 
From the exact Virasoro OPE, 
we know the leading operators in the global OPE \cite{Liendo:2012hy} are 
\be
\phi_{1,2}\times\phi_{1,2}=I+P_{1,3}\,\phi_{1,3}+P_T\,T+(P_3\,\phi_3+\dots),\label{OPEtrunP3}
\ee
where $\phi_3$ is a spin-2 primary with twist $\t_3=\D_{1,3}$. 
The absolute value of $P_T$ is larger than that of $P_3$, 
so it is consistent to omit $\phi_3$ and use \eqref{OPEtrun} as a crude approximation. 
In Figure \ref{fig-near-Ising}, we present the predictions of $\D_{1,3}$ from error minimizations 
using the truncated OPE \eqref{OPEtrun} with different $\D_{1,2}$. 
We notice that the $k$-dependence of the predictions is changed when $\D_{1,2}$ passes $1/8$, 
and the results are most stable at $\D_{1,2}=1/8$. 
We think these two phenomena are related to the behavior of $P_3$, 
the squared OPE coefficient of the subleading spin-2 operator $\phi_3$.
From the exact solution, 
the signs of $P_3$ are different for $\D_{1,2}<1/8$ and $\D_{1,2}>1/8$, 
and $P_3$ vanishes at $\D_{1,2}=1/8$. 
It seems the change in $k$-dependence is related to the different signs of $P_3$. 
In addition, as $P_3=0$ at $\D_{1,2}=1/8$, 
the hierarchy in the OPE coefficients is larger. 
Therefore, the truncated OPE \eqref{OPEtrun} becomes a better approximation around $\D_{1,2}=1/8$ 
and give more stable results. 

In principle, we can determine $\D_\s$ according to 
the stability of the predictions from error minimizations. 
We compute the predictions of ${\D_{1,3}}$ 
from $k=5,6,\dots,15$ with $\D_{1,2}=0.12,\,0.125,\,0.13$. 
The standard deviation of $\D_{1,3}$ with $\D_{1,2}=0.125$ is smaller than the other two cases, 
so this crude comparison gives: $\D_{1,2}=0.125(5)$. 
However, when $\D_{1,2}$ is very close to $0.125$, 
$|P_3|$ decreases \footnote{From the exact solution, $|P_3|\sim 10^{-4}$ at $\D_{1,2}=0.12$ and $\D_{1,2}=0.13$.} 
and $\phi_3$ is not the leading high-lying operator anymore. 
A precise determination of $\D_\s$ requires the introduction of other high-lying operators. 

We can consider a longer truncated OPE. 
For example, we can take into account the level-4 Virasoro descendants, 
then the number of free parameters grows significantly. 
We need to consider much larger $k$'s in the error function $\eta(k)$, 
which requires more sophisticated numerical methods. 
We leave this investigation for the future. 

Nevertheless, to test the error-minimization method, 
we can assume the spectral data are known and compute the OPE coefficients. 
In Table \ref{exact-long-OPE}, we summarize the results for the truncated OPE 
\be
\s\times\s=I+P_\epsilon\,\epsilon+P_T\,T+P_4\,\phi_4+P_5\,\phi_5+P_6\,\phi_6\,,\label{Ising-long-OPE}
\ee 
with the spectral data
\be
\D_\s=1/8,\, \D_\epsilon=1,\,
\{\D_4,\,l_4\}=\{4,\,0\},\,
\{\t_5,\,l_5\}=\{0,\,4\},\,
\{\t_6,\,l_6\}=\{1,\,4\}.
\ee
Since the spectral data are fixed, the predictions stabilize rapidly as we increase $k$. 
As expected, the predictions of $P_\epsilon,\, P_T$ are more accurate than those from the shorter OPE \eqref{OPEtrun}: 
the differences between the predictions and exact results are decreased by one order of magnitude.  
The results of $P_4,\,P_5,\,P_6$ are not very accurate due to their small magnitudes, 
but the orders of magnitudes are consistent with the exact values. 
To improve the accuracy of $P_4,\,P_5,\,P_6$, we need to consider more subleading operators, such as the level-6 Virasoro descendants. 
The minima of $\eta(k)$ with $k>5$ are of order $10^{-6}$, 
which are consistent with the magnitudes of the leading high-lying OPE coefficients. 

\begin{table}[h!]
\centering
 \begin{tabular}{||c ||c c c c c c || c ||} 
 \hline
 	$k$		&5		&  6 &7 & 8 & 9 & 10 &    exact\\ [0.5ex] 
 \hline
 \hline
$P_\epsilon$ 	& 0.2490 	& 0.2492	& 0.2492	& 0.2492 	& 0.2492 	& 0.2491 	& 0.25	\\\hline
$P_T$ 		& 0.0151	& 0.0152	& 0.0152	& 0.0152	& 0.0152	& 0.0152	&0.0156	\\\hline
$P_4$ 		&0.00018	&0.00019	&0.00019	&0.00019	&0.00019	&0.00019	&0.00024 \\\hline
$P_5$ 		&0.00009	&0.00011	&0.00011	&0.00011	&0.00011	&0.00011	&  0.00022	\\\hline
$P_6$ 		&0.00004	&0.00003	&0.00003	&0.00003	&0.00003	&0.00003	& 0.00002 	\\\hline
\end{tabular}
\caption{
The predictions from the error-minimization method of the squared OPE coefficients of the 2d Ising CFT 
and the corresponding exact values with a few significant figures. 
We use the truncated OPE \eqref{Ising-long-OPE} and assume the spectral data take the exact values.  
The predictions are slightly improved as we increase $k$.
}
\label{exact-long-OPE}
\end{table}

To obtain reasonable error bars of the predictions, we need to use more equations 
because the precision of the approximate crossing solutions is higher than the previous case \eqref{OPEtrun}. 
We will not discuss the error bars in this case. 

\section{Discussion}\label{discussion}
In this work, we study two aspects of conformal field theories:
 
The first part is about 4-point correlators around the fully crossing symmetric point $u=v=1$. 
Some general properties of the 4-point correlator of identical scalar operators are derived. 
We show that, around $u=v=1$, a generic 4-point correlator is close to the generalized free case. 
A new measure of deviations from the generalized free theory is proposed. 

In the second part, we present a new method for the conformal bootstrap with OPE truncations, 
which generalizes Gliozzi's method \cite{Gliozzi:2013ysa}. 
We propose to extract the low-lying CFT data by minimizing the OPE-truncation ``errors",  
i.e. the violations of crossing symmetry due to OPE truncations.
Geometrically, these error functions are the lengths of the vectors corresponding to the truncated OPEs. 
The minimum of an error function also gives an estimate of the OPE coefficient of the leading high-lying operator omitted in the truncated OPE. 
We apply the error-minimization method to the 2d Ising CFT. 
We show the effectiveness of this new method and 
discuss how the predictions are influenced by the hierarchy in the OPE coefficients. 
Although the OPEs are severely truncated, 
the results of the 2d Ising CFT are fairly accurate and stable due to 
the decoupling of the subleading spin-2 primary operator. 
In addition, the error-minimization method can be used to extract OPE coefficients 
when the low-lying spectrum is known, for example in the extremal functional method \cite{Poland:2010wg,ElShowk:2012hu}. 
\\

Let us discuss some possible directions for future research. 
 
For the first part, we can generalize the results about the fully crossing symmetric point. 
1) There are fully crossing-symmetric solutions with closed form expressions 
controlled by the other independent parameters. 
They deserve a more systematic investigation.  
2) For the correlators of identical scalars, the second crossing equation is satisfied 
by each conformal block associated with a primary of even spins. 
However, for a generic correlator, the situation can be different and 
the general constraints from the second crossing equations may contain useful information. 
These two studies could lead to constructive results for the inverse conformal bootstrap \cite{Li:2017agi}. 

For the second part, there are several interesting questions concerning the error-minimization method:
\begin{itemize}
\item 
By using efficient numerical techniques, we can study truncated OPEs with more operators. 
Then we have access to the refined structure of OPEs and can study more complex theories, 
such as 2d irrational CFTs 
\footnote{The numerical modular bootstrap \cite{Collier:2016cls,Bae:2017kcl,Dyer:2017rul}   can also be carried out 
by the error-minimization method 
with minor adaptations 
if the convergence of character expansions is rapid. } 
and higher dimensional CFTs. 
\item
We may improve the method by considering more sophisticated error functions. 
In \eqref{error}, we use the equations from the Taylor expansion around a crossing symmetric point 
and assign the same weight to each equation. 
The way we choose equations and weights is not well justified. 
The equations from higher derivatives should be more sensitive to high-lying operators, 
so the predictions for a fixed truncated OPE could become unstable when $k$ in \eqref{error} is extremely large. 
The choice of weights is purely for simplicity. 
It seems more natural to consider the crossing equations 
evaluated at different points in the rapidly convergent regime \cite{Hogervorst:2013sma, Echeverri:2016ztu} 
and use them to define the error functions with appropriate weights. 
It will also be very interesting to implement the error-minimization method in an analytic manner. 
 
\item 
We want to know if rigorous bounds can be obtained  
using the triangle inequality in normed vector spaces, 
based on the small magnitudes of high-lying OPE coefficients. 
The role of vector lengths here will be in parallel to that of vector directions in the bound method based on unitarity. 
\end{itemize}

\section{Acknowledgements}
I would like to thank Euihun Joung and Junchen Rong for stimulating discussions. 
I thank Slava Rychkov for constructive comments. 
I also want to thank Connor Behan for helpful correspondence. 
This work was supported by the National Research Foundation of Korea through the grant NRF-2014R1A6A3A04056670.

\end{document}